\documentclass[12pt]{article}
\usepackage[dvips]{graphicx}
\usepackage{amssymb,amsfonts,amsmath,latexsym}
\usepackage{epsfig, graphicx}

\usepackage{amstext}
\usepackage{epsfig, graphicx}
\usepackage{epsfig}
\usepackage{latexsym}
\usepackage{amstext}
\usepackage{amssymb}
\usepackage{graphics}
\usepackage{color}

\newcommand{\p}{\partial}
\newcommand{\ds}{\displaystyle}
\newcommand{\beq}{\begin{eqnarray}}
\newcommand{\beqq}{\begin{eqnarray*}}
\newcommand{\eeq}{\end{eqnarray}}
\newcommand{\eeqq}{\end{eqnarray*}}

\newcommand{\x}{\mbox{\boldmath$x$}}

\newcommand{\y}{\mbox{\boldmath$y$}}

\newcommand{\n}{\mbox{\boldmath$n$}}
\newcommand{\X}{\mbox{\boldmath$X$}}
\newcommand{\w}{\mbox{\boldmath$w$}}

\newcommand{\Bb}{\mbox{\boldmath$b$}}

\definecolor{red}{rgb}{1,0,0}

\def\ds#1{\displaystyle{#1}}

\begin{document}
%%%%%%%%%%%%%%%%%%%%%%%%%%%%%%%%%%%%%%%%%%%%%%%%%%%%%%%%%%%%%%%%
\title{Residence times of receptors in dendritic spines analyzed by simulations in empirical domains}
%%%%%%%%%%%%%%%%%%%%%%%%%%%%%%%%%%%%%%%%%%%%%%%%%%%%%%%%%%%%%%%%
\author{ N. Hoze$^{1*}$ and D. Holcman$^{1}$}

\maketitle
\let\thefootnote\relax\footnotetext{$^1$Group of Computational Biology and Applied Mathematics, IBENS, INSERM-1024, Ecole Normale Sup\'erieure, 46 rue d'Ulm 75005 Paris, France.}

\begin{abstract}
Analysis of high-density superresolution imaging of receptors reveal the organization of dendrites at the nano-scale resolution. We present here simulations in empirical live cell images, which allows converting local information extracted from short range trajectories into simulations of long range trajectories. Based on these empirical simulations, we compute the residence time of an AMPA receptor (AMPAR) in dendritic spines that accounts for receptors local interactions and  geometrical organization. We report here that depending on the type of the spine, the residence time varies from one to five minutes. Moreover, we show that there exists transient organized structures, previously described as potential wells  that can regulate the trafficking of AMPARs to dendritic spine:
%as long as the well persisted, AMPARs could not enter the spine and after it disappeared a {large number of receptors} could enter.
The simulation method developed here and the results suggest that AMPAR trafficking is regulated by transient structures.
\end{abstract}

{\bf keywords: AMPAR | trafficking | first passage times| residence times | dendritic spines |  Stochastic analysis and simulations of trajectories | Diffusion | Molecular Interactions}

%%%%%%%%%%%%%%%%%%%%%%%%%%%%%%%%%%%%%%%%%%%%%%%%%%%%%%%
%\section*{Introduction}\label{ sec:intr}
%%%%%%%%%%%%%%%%%%%%%%%%%%%%%%%%%%%%%%%%%%%%%%%%%%%%%%%
Receptor trafficking has been identified as a key feature of synaptic transmission and plasticity \cite{Bredt1,Tomita,Schnell2002,Malinow2,Malinow3}. Yet, the mode of trafficking remains unclear: classical single particle tracking revealed that after receptors are inserted in the plasma membrane of a neuron, their motion can either be free or confined Brownian motion \cite{Choquet-triller}. Recently,  superresolution light optical microscopy techniques for \textit{in vivo} data \cite{Choquet2,Saxton2008,HozePNAS,DeepakJNS2013,Masson} have allowed monitoring a large number of  trajectories at the single-molecule level and at nanometer resolution. High confined density regions are generated by large potential wells (100s of nanometers) that sequester receptors \cite{HozePNAS,Masson}. These potential wells were anticipated theoretically and use to describe receptor confinement in \cite{holcman2004}(p.995) and \cite{taflia}. In addition, fluctuations in the apparent diffusion coefficient reflect  the changes in the local density of obstacles \cite{Saxton2007,Saxton-95,Hoze2}. Classically, cell membranes are organized in local microdomains \cite{Sheetz,KusumiReview} characterized by morphological and functional specificities. In neurons, prominent microdomains include dendritic spines and synapses, which play a major role in neuronal communication.\\
Because receptor density at a synapse determines the synaptic strength \cite{Bredt1,Malinow2}, it is essential to estimate their numbers and residence time inside a synapse. It was very difficult to access experimentally using FRAP (Fluorescent Recovery after Photobleaching) or stable Quantum Dots, the residence time of receptors at synapses or at the Post-Synaptic Density (PSD), due to the small size and the exact delimitation of these microdomains. The number of receptors has been estimated using coarse-grained models of receptor trafficking \cite{Holcman-triller,Bressloff0} in idealized spine geometries. \\
Our goal here is to compute the residence time of AMPA receptors in dendritic spines using short AMPA receptor trajectories, much shorter than the total residence time. We develop here a  novel approach to compute from many short trajectories the global mean residence time in micrometer domains. This time depends singulary on geometrical parameters such as the neck radius for dendritic spines, as estimated in \cite{diffusion-laws,Holcman-schuss2013}. The present analysis relies on simulations in empirical live cell images, which allow converting local biophysical information, extracted from a large amount of short-range trajectories, into numerical simulations of long-range trajectories. The method of extracting local biophysical properties uses the classical Smoluchowski approximation of the Langevin's equation. We simulate from the extracted stochastic equation, long trajectories for which the diffusion tensor and the local force are directly obtained from empirical data. Furthermore, to emphasize the applicability of our method, we show that AMPA receptor trafficking are affected by stable and/or transient potential wells. For example, we find that the presence of a potential well at the base of a dendritic spine can prevent receptors to enter into a dendritic spine and as soon as the potential well disappears, a large amount of AMPARs can enter through a dendritic spine neck up to the head. We propose that potential wells are structures that also modify the inward flux of receptors, a process that can regulate synaptic plasticity and should be added to the classical framework of lateral diffusion \cite{Saxton-97}.
%%%%%%%%%%%%%%%%%%%%%%%%%%%%%%%%%%%%%%%%%%%%%%%%%%%%%%%%%%%%%%%%%%%%%%%%%%%%%%%%%%%%%%%%%%%%%%%%%%%%%%%%%%%%
\section{Methods and theory}
%%%%%%%%%%%%%%%%%%%%%%%%%%%%%%%%%%%%%%%%%%%%%%%%%%%%%%%%%%%%%%%%%%%%%%%%%%%%%%%%%%%%%%%%%%%%%%%%%%%%%%%%%%%%
Before presenting the method of the stochastic simulations in empirical domains, we recall the stochastic equations that describe the motion of receptors.

\subsection{The Langevin model for the neuronal receptors stochastic motion}
The residence time of a stochastic receptor inside a neuronal microdomain cannot be directly obtained by averaging over short recorded trajectories, because they do not necessarily span the entire neuronal space. Moreover, longer trajectories obtained by other techniques (such as quantum dot) only provide a partial sampling of the space, which cannot be used to compute any probability of transition between different regions. Because there is no direct method to estimate the residence time of a receptor in a microdomain such as a dendritic spine, we now present a simulation approach, where the local biophysical properties are obtained from a large sampling of single particle data.\\
The physical model of a receptor motion on a homogenous surface is the Smoluchowski's limit of the Langevin equation. For a receptor moving on a two-dimensional surface, containing many impenetrable obstacles, the dynamics is generated by a diffusion coefficient $D$ and a field of force $-\nabla U$
\beq\label{stochlocal0}
\dot{\X}=-\frac{\nabla U(\X)}{\gamma} +\sqrt{2D} \dot{\w},
\eeq
where $\w$ is the standard white noise. The field of force can also represent locally a drift. A zero flux boundary condition should be given on all obstacles, which leads cannot be simply expressed in the stochastic equation.  Thus, to account for the local crowding organization, equation \ref{stochlocal0} is usually coarse-grained (see \cite{Hoze2}) as
\beq\label{stochlocal01}
\dot{\X}=-\frac{\nabla U(\X)}{\gamma} +\sqrt{2D_e(\X)} \dot{\w},
\eeq
where the effective diffusion coefficient is $D_e(\X)$ and can be computed in some cases, as a function of the intrinsic diffusion D \cite{Hoze2}, density and geometry of obstacles. The effective diffusion coefficient can depend on the location $\X$, while the friction coefficient $\gamma$ remains a constant, as large obstacles contributing in slowing down the effective diffusion, should not affect the local physical properties of the diffusing particle. \\
By using an ensemble of trajectories, it is possible to invert equation \eqref{stochlocal0} and extract the local drift and the effective diffusion coefficient.  The reconstruction of the field of forces and the diffusion coefficient is given by the formulas \cite{book,DSP,OPT}
 \beq
 \label{eq:force}
-\frac{\nabla U(\X)}{\gamma}=\lim_{\Delta t \rightarrow 0} \frac{\langle\X(t+\Delta
t)-\X(t)\,|\,\X(t)=\X\rangle}{\Delta t},
\eeq
where $\langle\cdot\rangle$ represents the average over the trajectories passing through point $\X$ at time $t$. The inversion procedure requires combining several independent trajectories passing through each point of the neuronal surface. Similarly, the two-dimensional membrane diffusion tensor is given by
 \beq \label{eq:diff}
2{\bf D^{ij}}(\X)=\lim_{\Delta t \rightarrow 0} \frac{\langle (\X(t+\Delta
t)-\X(t))^i(\X(t+\Delta
t)-\X(t))^j,|\X(t)=\X\rangle}{\Delta t},
 \eeq
For completeness, we shall now explain how the drift and the diffusion coefficient are extracted from empirical data.
%%%%%%%%%%%%%%%%%%%%%%%%%%%%%%%%%%%%%%%%%%%%%%%%%%%%%%
%\subsection{Reconstruction of trajectories}
\subsection{Extraction of the local dynamics using non-parametric estimators}
\begin{figure}[ht!]
\begin{center}
\includegraphics[scale=0.75]{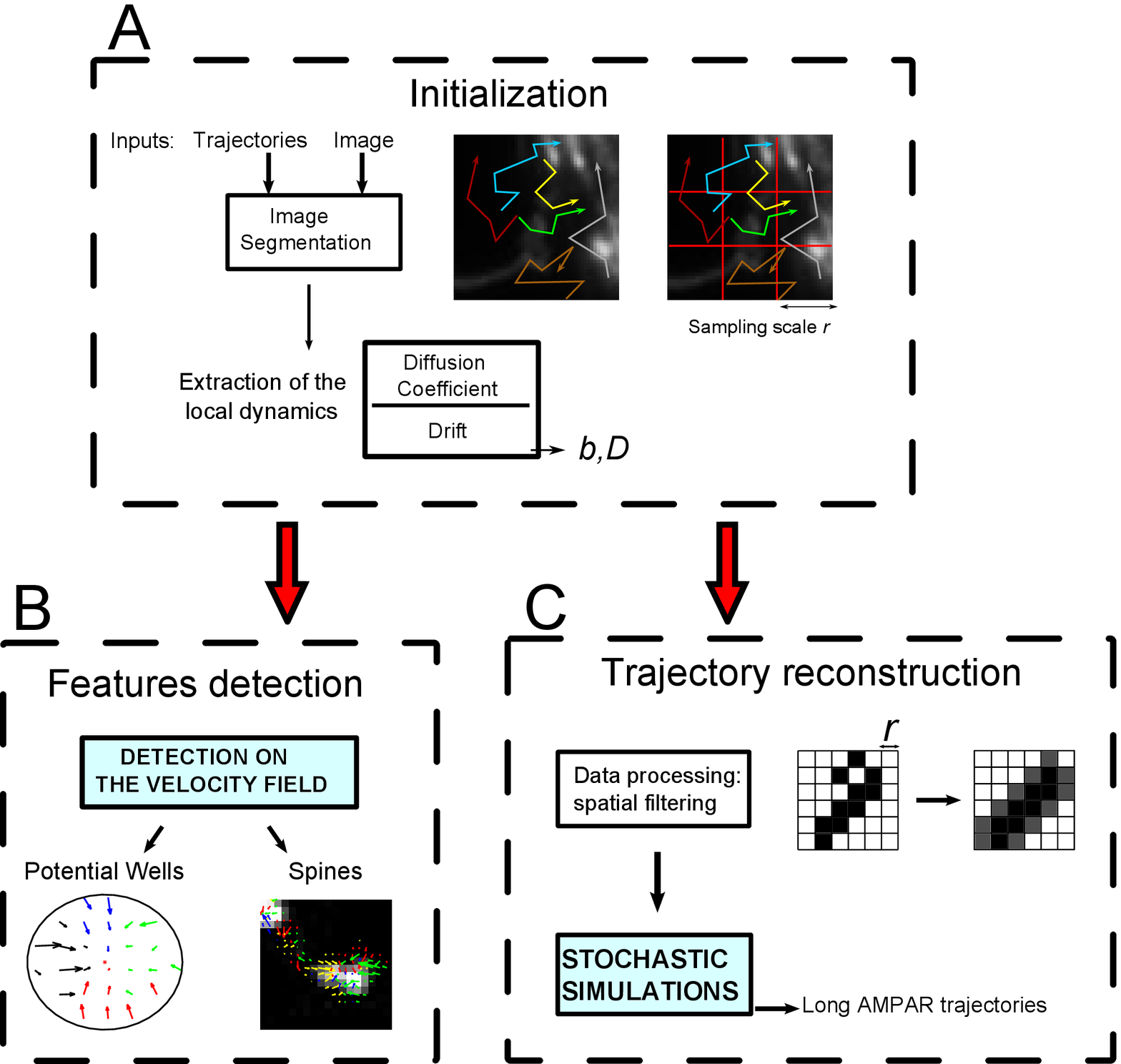}
\caption{ \textbf{ Schematic representation of the reconstruction algorithm.} (\textbf{A}) AMPAR trajectories represented on the confocal image of a neuron. We decompose the image into squares of length $r$. The drift $\Bb$ and diffusion $D$ components are computed by averaging the mean velocity and mean squared differences for the trajectories passing on each square. (\textbf{B}) The velocity map allows detecting specific features on the neuron, such as potential wells and properties of dendritic spines. (\textbf{C}) Sequential spatial filters (see Methods section for details) requires simulating AMPAR trajectories and recomputing local values of the drift and diffusion coefficient. Trajectories are simulated using an Euler's scheme for the stochastic equation $\dot \X_t=\Bb(\X_t)+\sqrt{2D(\X_t)} \dot \w(t)$. Parameters are extracted from empirical trajectories. Long trajectories can now be generated on the empirical domain.}
\label{algo}
\end{center}
\end{figure}

%%%%%%%%%%%%%%%%%%%%%%%%%%%%%%%%%%%%%%%%%%%%%%%%%%%%%%
We use a dataset consisting of a large amount of short trajectories (5 points in average sampled every 50 ms) of AMPA receptors moving on neurons, described in \cite{HozePNAS}. These data consisted of a massive amount of trajectories generated by the sptPALM method applied to Hippocampal AMPARs  cultured neurons \cite{Heine2008}.\\
To extract the local receptor dynamics, we use the ensemble of points where the amount of trajectories falling into the neighborhood of each point was sufficient (around 200). These short trajectories cover only a very small fraction of the space and cannot be used directly to obtain estimate about transition properties between any given region of interest. \\
We first reconstruct the diffusion coefficient and the drift according to the method described in \cite{HozePNAS} leading to the local drift  $\Bb(\X)$ and the diffusion tensor $D(\X)$ at position $\X$ (the value is given at a given resolution, which is the size of the squares $S(\X_c,r)$ of side $r$ and center at $\X_c$ (Fig. \ref{algo}A)). We use the empirical approximation of relations \eqref{eq:force}-\eqref{eq:diff}:

For a square $S(\x,r)$ centered at $\X$ of side $r$, when there are $N(\X,r)$ points of sampled trajectories $\y_1,...,\y_N$ falling into $S(\X,r)$, such that $\y_1=X_{i_1}(t_1),... \y_N=X_{i_N}(t_N)$, then we approximate the drift $\Bb(\X,r)$ at position $\X$ by is classical empirical sum
\beq
\Bb(\X,r) \approx \frac{1}{N(\X,r) }\sum_{k=1}^{N(\X,r)}
\frac{\X_{i_k}(t_k+\Delta t)-\X_{i_k}(t_k)}{\Delta t}.
\label{eqdrift}
\eeq
As $r$ goes to zero and $N$ is fixed, the quality of the approximation increases. This is an optimal unbiased estimator.

%The second feature to extract from the AMPAR trajectories is the
%random part of stochastic equation \eqref{stochlocal}. In a given
%system of coordinates, the characterization of the tensor
%\cite{Karlin} is
%\beq
%2D\sigma^{ij}(\x)=\lim_{\Delta t \rightarrow 0}
%E\left(\frac{\left(\X^i(t+\Delta t)-\X^i(t)\right)(\X^j(t+\Delta
%t)-\X^j(t))}{\Delta t}|\X(t)=\x\right).
%\eeq
%where $D$ is the intrinsic diffusion constant of the receptor. Using
%our trajectories, following the same procedure as the one developed
%for the drift, in a square $S(\x,r)$ centered at $\x$ of side $r$,
%when there are $N(\x,r)$ points of the trajectories
%$\y_1=X_{i_1}(t),...,\y_N=X_{i_N}(t)$,
Similarly, we approximate the tensor $\sigma^{ij}(\X,r)$ at position $\X$ and resolution $r$ by
\beq
2D \sigma^{ij}(\x)\approx \frac{1}{N(\X,r)}\sum_{k=1}^{N(\X,r)}
\frac{(\X^i(t+\Delta t)-\X^i(t)(\X^j(t+\Delta t)-\X^j(t)))}{\Delta
t}.
\label{tensor}
\eeq
$\sigma^{ij}(\x)$ contains information about the local organization of the neuronal surface. In practice, the diffusion is isotropic and the diffusion tensor $\sigma^{ij}$ is proportional to the identity matrix.  Other parametric estimators are presented in \cite{Masson}.

%%%%%%%%%%%%%%%%%%%%%%%%%%%%%%%%%%%%%%%%%%%%%%%%%%%%%%
\subsection{Imaging analysis, spatial filtering and numerical discretization}
\begin{figure}[ht!]
\begin{center}
\includegraphics[scale=0.7]{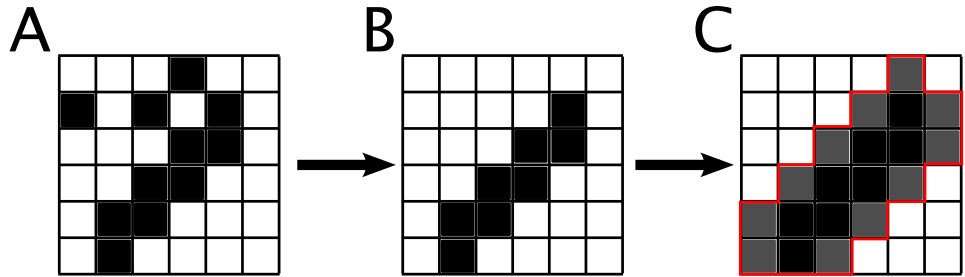}
\caption{ {\bf Example of spatial filtering for trajectory simulation.}
A dendritic spine is pixelized: (\textbf{A}) image before reconstruction,
(\textbf{B}) after suppression of the regions with less than 15 trajectories and
of isolated regions and (\textbf{C}) after application of the low-pass
filter (eq.\ref{lowpassfilter}). The region $\Omega$ is in black and gray, $\Omega_0$ in white, and the boundary $\p \Omega$ (in red).}
\label{suppfig3}
\end{center}
\end{figure}

%%%%%%%%%%%%%%%%%%%%%%%%%%%%%%%%%%%%%%%%%%%%%%%%%%%%%%
Because the image resolution was lower than the one of the trajectories, we found several artifacts (due to the pixelization) that we resolve here using a spatial filtering. We will generate trajectories inside live microscopy empirical images using the stochastic approximation of equation \eqref{stochlocal0}
\beq
\dot \X_c =\Bb(\X_c) +\sqrt{2D(\X_c)} \dot \w,
\label{eqmotion}
\eeq
where $\Bb(\X_c)$ and $D(\X_c)$ are the piecewise constant value of the drift and the diffusion coefficient at any point $\X \in S(\X_c,r)$. This discretization procedure of the space (neuronal dendrite) in small squares $S(\X_c,r)$ generates local discontinuity and artificially non-connected regions, which we overcome using a sequence of spatial filters. A schematic summary of the algorithm is described in Fig. \ref{algo}.\\
{\bf \noindent Construction of the spatial filter}: small size of microdomains such as dendritic spines is reflected by the small amount of pixels dedicated to represent them (around 20 and 30). This small number creates an additional difficulty to simulate stochastic trajectories: indeed in a pixelized image, some pixels intersect the region $\Omega_0$ not covered by experimental trajectories,at corners and in that case, switching from one pixel to another by a stochastic trajectory corresponds to a very unlikely jump, as it is a rare event. Consequently, the residence time falls into the class of escape problem \cite{Holcman-schuss2013,DSP}. In that specific case, such trajectory is artificially restricted and the confinement time has no biological relevance. To avoid this confinement, we suppressed pixels surrounded by the ones in the ensemble $\Omega_0$ or by those touching $\Omega$ (the pixel ensemble) at corners (see Fig. \ref{suppfig3}). This procedure consists in eliminating pixels as follows: A pixel of size $r$ and coordinates $(i,j)$ is suppressed when the four neighbors are in the ensemble $\Omega_0$. In that case, we replace the diffusion coefficient at the pixel by its local average
\beq
\left\{\begin{array}{lll}
D_{i,j}&\neq& 0\\ \\
D_{i+1,j}&=&D_{i-1,j}=D_{i,j-1}=D_{i,j+1}=0,
\end{array}\right.
\label{eq 1}
 \eeq
where $D_{i,j}$ (resp. $\Bb_{i,j}$) is the discrete value of the diffusion tensor $D(\X)$ (resp. drift $\Bb(\X)$) in the pixel of two-dimensional coordinates $(i,j)$. This procedure corresponds to running a low-pass filter that smooths out the diffusion coefficients and the drift term. Indeed, the procedure is a weighted average on the square centered at $(i,j)$ and on the four adjacent squares given by
\beq
D'_{i,j}=\frac{1}{2}D_{i,j}+\frac{1}{8}(D_{i-1,j}+D_{i+1,j}+D_{i,j-1}+D_{i,j+1}).
\label{lowpassfilter}
\eeq
To simulate trajectories at each point of the dendrite image, we discretized equation \ref{eqmotion} using the Euler's scheme:
\beq \label{stochlocaldiscrete}
\X(t+\Delta t)=\X(t)+\Bb_r(\X)\Delta t+\sqrt{2D_r(\X)\Delta t}{\bf
\eta},
\eeq
where ${\bf \eta}$ is a two-dimensional normalized Gaussian variable and $\Delta t$ is the simulation time-step.  In region $\Omega_0$ not covered by experimental trajectories, we took the condition $\sigma(\X)=\Bb(\X)=0$. The simulated trajectories are reflected on the boundary $\p \Omega$ of the pixel ensemble. We also added to the uncovered region the  squares where the sampling was insufficient ($<15$ points).

%%%%%%%%%%%%%%%%%%%%%%%%%%%%%%%%%%%%%%%%%%%%%%%%%%%%%%
\subsection*{Reflecting procedure at a boundary and choice of the ideal time step}
%%%%%%%%%%%%%%%%%%%%%%%%%%%%%%%%%%%%%%%%%%%%%%%%%%%%%%
We shall describe here the procedure of reflection of a trajectory at the boundary. For two consecutive simulated points $\X(t)$ and $\X(t+\Delta t)$, generated by equation \eqref{stochlocaldiscrete}, where $\X(t+\Delta t)$ has crossed the boundary $\p \Omega$, the zero flux boundary condition leads to
the Snell-Descartes reflection condition,  where the reflection point $\X'(t+\Delta t)$ is computed from $\X(t+\Delta t)$ by
\beq
(\X'(t+\Delta t)-P).\n=-(\X(t+\Delta t)-P).\n ,
\eeq
where $\n$ is the normal to $\p \Omega$ at the boundary point $P$. The tangent component is unchanged.\\
The time step for the simulation associated with equation \eqref{stochlocaldiscrete} is chosen following the following empirical criteria:  a simulated trajectory spends at least a mean time of $5 \Delta t$ in an elementary square  $S(\X,r)$.  From equation \eqref{stochlocaldiscrete}, using the condition
\beq
||\X(t+5\Delta t)- \X(t)|| \leq r.
\eeq
We obtain the following condition on  $\Delta t$
\beq
25 |\Bb(\X)|^2\Delta t^2 +10 D(\X)\Delta t   \leq r^2.
 \eeq
Thus we obtain  for all squares $S(\X_k,r),$
\beq
\Delta t(\X_k)=\frac{ D(\X)}{5|\Bb(\X)|^2}\left(-1+\sqrt{1+\left( \frac{r |\Bb(\X)| }{D(\X)}\right)^2} \right).
 \eeq
Finally, we chose
\beq
\Delta t= \min_{\X_k \ds \in \Omega} \Delta t(\X_k).
 \eeq

We can now apply the method describe above to estimate the residence time of reconstructed stochastic trajectories in dendritic spines by simulating long trajectories from equation \eqref{stochlocal0} and its approximation \eqref{eqmotion}, starting inside the spine head until they reach the dendritic shaft.

%%%%%%%%%%%%%%%%%%%%%%%%%%%%%%%%%%%%%%%%%%%%%%%%%%%%%%%%%%%%%%%%%%%%%%%%%%%%%%%%%%%%%%%%%%%%%%%%%%%%%%%%%%%%
\section{Results}
%%%%%%%%%%%%%%%%%%%%%%%%%%%%%%%%%%%%%%%%%%%%%%%%%%%%%%%%%%%%%%%%%%%%%%%%%%%%%%%%%%%%%%%%%%%%%%%%%%%%%%%%%%%%

%%%%%%%%%%%%%%%%%%%%%%%%%%%%%%%%%%%%%%%%%%%%%%%%%%%%%%
\subsection{Residence time of AMPARs in dendritic spines}
\begin{figure}[ht!]
\begin{center}
\includegraphics[scale=0.7]{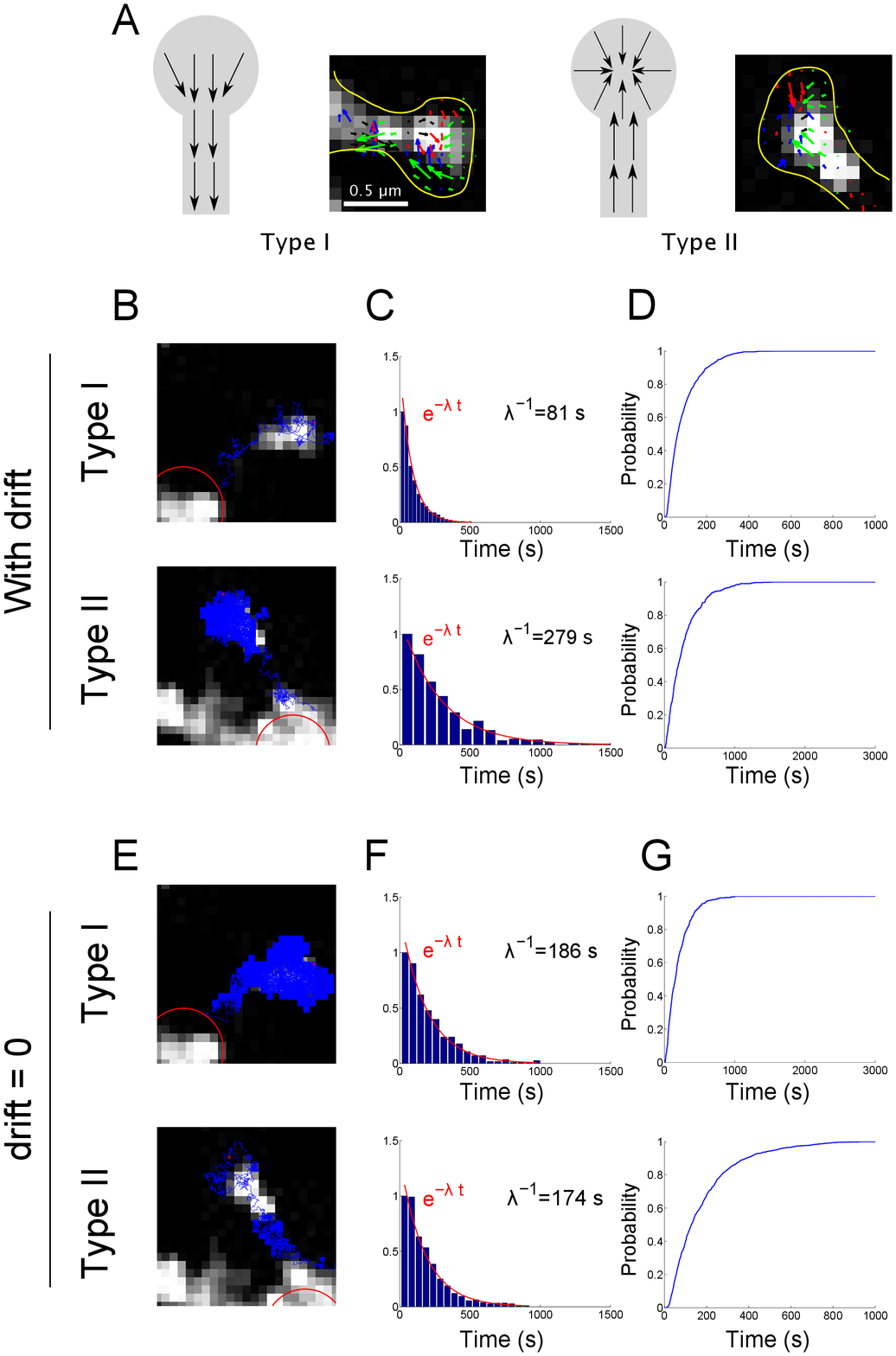}
\caption{ {\bf Residence time of a receptor in  a dendritic spine  obtained by numerical simulations.} (\textbf{A}) Two types of spines:  Type I (resp. Type II) are characterized by an outward drift (resp. inward drift). (\textbf{B}) Brownian simulations of trajectories exiting a Type I and a Type II spine, using equation \eqref{eqmotion}. Parameters are extracted from single particle trajectories. A trajectory starting at the red spot ends when it hits the target area (red circle).  (\textbf{C}) Histograms of the residence time showing an asymmetric distribution associated to the different spine types.  The values of the exponential decays with rate $\lambda$ are indicated. (\textbf{D}) Cumulative distribution function of the first time to exit the spine head.  1,000 simulations were performed for each spine. In (\textbf{E,F,G}), the drift in equation \eqref{eqmotion} is set to zero.}
\label{figsimulations}
\end{center}
\end{figure}

%%%%%%%%%%%%%%%%%%%%%%%%%%%%%%%%%%%%%%%%%%%%%%%%%%%%%%
To estimate the residence time of AMPAR, we applied the procedure described in the Methods section. We first estimated the drift and diffusion coefficient and generated  trajectories from the stochastic equation \eqref{stochlocal0}. We selected dendritic spines geometry, applied the filtering procedure and then simulated long trajectories from the coarse-grained stochastic equation \eqref{eqmotion}.\\
We applied this procedure to Type I and Type II spines (Fig. \ref{figsimulations}A), where the drift in the neck is inward and outward respectively. We found that the residence time depends on the spine geometry and the nature of the spine Fig. \ref{figsimulations}B-D: For Type I, we found a residence time of 81 s, while it is of the order of 279 s for Type II.\\
To assess the contribution of the drift in the residence time, we performed simulations of pure Brownian trajectories  by setting the drift component in the neck and in the PSD to zero. We found (Fig. \ref{figsimulations}E-G) that the mean residence time was around 180 s for both spine types. At this stage, we conclude that the excess of time for type II is due to the inward drift inside the neck and the internal potential well at the PSD.
%%%%%%%%%%%%%%%%%%%%%%%%%%%%%%%%%%%%%%%%%%%%%%%%%%%%%%%%%%%%%%%%%%%%%%%%%%%%%%%%%%%%%%%%%%%%%%%%%%%%%%%%%%%%
\subsection{Transient versus stable neuronal nanodomains}
\begin{figure*}[ht!]
\begin{center}
\includegraphics[scale=.9]{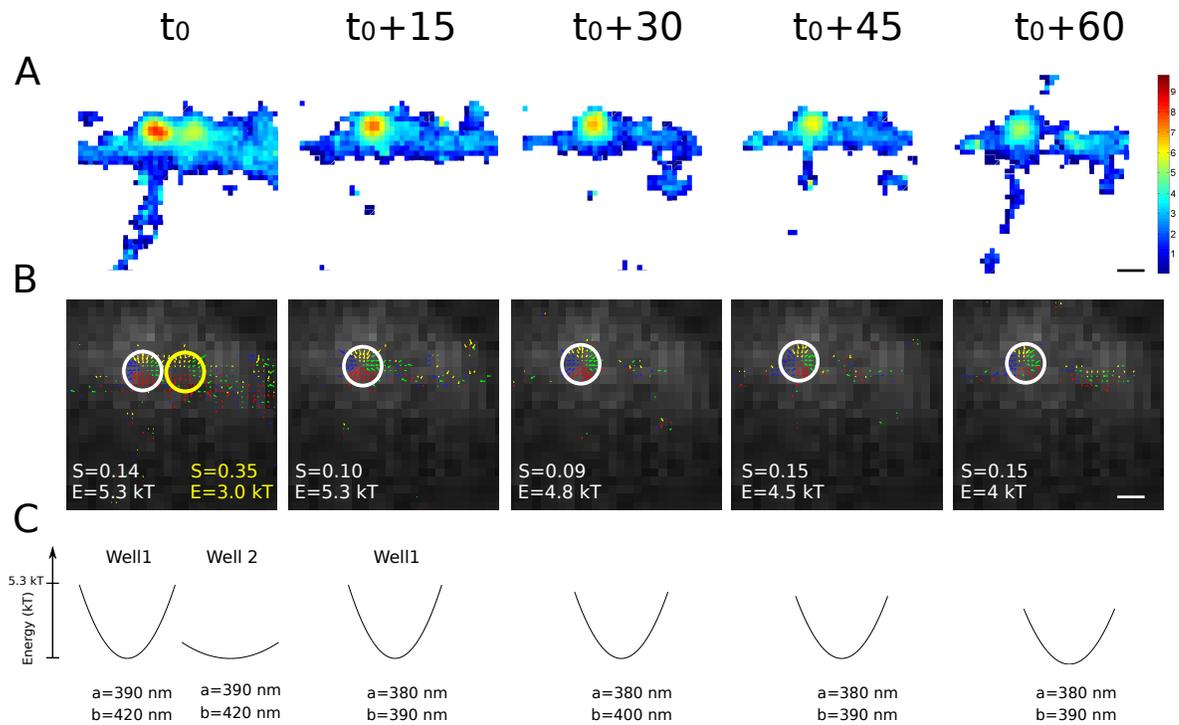}
\caption{ { \bf Time lapse analysis of two potential wells.}
(\textbf{A}) Density of points obtained from AMPAR trajectories.
(\textbf{B}) Drift map revealing two types of potential wells: a very stable one that persists over one hour and a transient one that disappears in 30 minutes. (\textbf{C}) Shape (base and depth) over time of the two potential wells described in (\textbf{B}). Scale bars: 500 nm.}
%(\textbf{H})  Approximation of the potentials by optimal fitting of generic parabolic wells.}
\label{fig1}
\end{center}
\end{figure*}

%%%%%%%%%%%%%%%%%%%%%%%%%%%%%%%%%%%%%%%%%%%%%%%%%%%%%%%%%%%%%%%%%%%%%%%%%%%%%%%%%%%%%%%%%%%%%%%%%%%%%%%%%%%%
 Receptor trajectories can be attracted by long range potential wells. The wells are of few hundreds of nanometer in size and their interaction energy is of few kT (less than 6 or between 8 to 10). To elucidate their nature, we decided to estimate the characteristics of the potential wells over time. We report now the characteristics of three potential wells. We observed over time two neighboring wells (Fig. \ref{fig1}A,B):  the first one is very stable and lasts more than one hour with small decay in energy, although there was some local spatial reorganization (Fig. \ref{fig1}C). In contrast, the second neighboring one is very transient (Fig. \ref{fig1}A,B). It is only located a few hundreds nanometers (700 nm) apart from the first one but it disappears in about 30 minutes. The changes are drastic both in size and energy. We conclude that potential wells can be very transient, but the underlying molecular structure is quite localized: when a potential well disappears, it does not affect another one located hundreds of nanometers apart.
%%%%%%%%%%%%%%%%%%%%%%%%%%%%%%%%%%%%%%%%%%%%%%%%%%%%%%%%%%%%%%%%%%%%%%%%%%%%%%%%%%%%%%%%%%%%%%%%%%%%%%%%%%%%
\subsection{Regulation of AMPAR entry into dendritic spines}
\begin{figure}[ht!]
\begin{center}
\includegraphics[scale=.9]{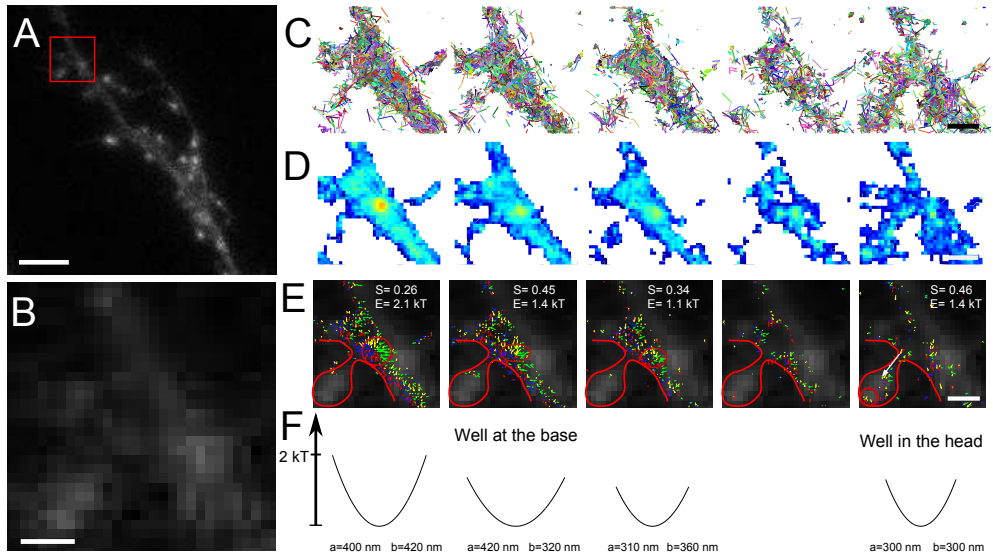}%{FigureTimeLapse-4.pdf}
\caption{ {\bf A potential well on the dendritic shaft prevents receptors from entering into a dendritic spine.}
Trajectories were acquired during one minute every fifteen minutes.
(\textbf{A}) Confocal image of the dendrite from a cultured hippocampal neuron.
(\textbf{B}) Magnification of the region outlined by the red square in (\textbf{A}).
(\textbf{C}) AMPAR trajectories between the dendritic shaft and a single spine.
(\textbf{D}) Density of points obtained from trajectories of \textbf{C}.
(\textbf{E}) As long as the potential well at the base of the spine (red circle) persists, no AMPAR trajectories can enter into the spine. After 45 minutes, the potential disappears and a large amount of trajectories can be found inside the spine head, maintained by a potential well. (\textbf{F}) Characteristics over time of the potential well  at the base of the dendritic spine.
 Scale bars, \textbf{A}: 5$\mu$m, \textbf{B}, \textbf{C}, \textbf{D}, \textbf{E}: 1 $\mu$m. }
\label{figtimelapse}
\end{center}
\end{figure}

%%%%%%%%%%%%%%%%%%%%%%%%%%%%%%%%%%%%%%%%%%%%%%%%%%%%%%%%%%%%%%%%%%%%%%%%%%%%%%%%%%%%%%%%%%%%%%%%%%%%%%%%%%%%
Although the role of stable potential wells at synapses is to recruit and maintain receptors,
the function of transient potential wells is still unclear. To address their possible role, we monitored over time a potential well located at the base of a dendritic spine (Fig. \ref{figtimelapse}A-B). The time lapse analysis of trajectories  at the junction between the spine and dendritic shaft (Fig. \ref{figtimelapse}C-D) revealed a region of high density at the base of the spine, which we identified as a stable potential over a period of 30 minutes. The characteristics of the well  are given in Fig. \ref{figtimelapse}E-F. Interestingly, as long as the potential well was deep enough (large energy barrier), dendritic trajectories could not penetrate from the dendrite to the spine. However, when the potential disappeared, we observed a large receptor influx inside the spine up to the spine head. This result suggests that the role of the wells is not only to retain receptors at the post-synaptic density as already shown in \cite{HozePNAS}, but for the wells located in the dendrite, to additionally prevent receptors from entering into microdomains such as dendritic spines.

%%%%%%%%%%%%%%%%%%%%%%%%%%%%%%%%%%%%%%%%%%%%%%%%%%%%%%
\section{Discussion}
%%%%%%%%%%%%%%%%%%%%%%%%%%%%%%%%%%%%%%%%%%%%%%%%%%%%%%
Constructing long-range trajectories based on local biophysical properties is now possible and the procedure consists in finding the drift and the diffusion tensor from massive trajectories filling the available space. We used here the Langevin's equation to extract the biophysical properties from data and then generated long-range trajectories that would resemble real and ideal trajectories. Trajectories are generated in a confocal image of a neuron that has been discretized. This discretization can create a distortion of the motion by making local cavities, which induce narrow passages that would retain receptors for arbitrarily long times \cite{Holcman-schuss2013,Direstrait2012}. We thus filter these artificial cavities and generated trajectories with the parameters obtained from the original data. The present method described here is quite general. It allows computing the residence time of a trajectory from any domain of interest.

%%%%%%%%%%%%%%%%%%%%%%%%%%%%%%%%%%%%%%%%%%%%%%%%%%%%%%
\subsection{Residence time of AMPAR in dendritic spines}
%%%%%%%%%%%%%%%%%%%%%%%%%%%%%%%%%%%%%%%%%%%%%%%%%
We estimated here the residence time of AMPARs in dendritic spines from superresolution data. This data allows   computing the residence time of receptors in single dendritic spines,   at the molecular resolution (single particle tracking), giving a higher   precision compared to FRAP experiments. Indeed FRAP provides information at a population level for the fraction of mobile receptors. The time scale of diffusion  in spine heads in the range 100-222 s \cite{Ashby}, which is comparable with the order reported here. Indeed, we noted that the mean time $t_{1/2}$ used to fit as a single exponential the FRAP data is precisely the mean residence time of a receptor in the associated structure \cite{Holcman-schuss2013}. The large disparity in the residence time of receptors in the spine depends not only on the geometry (see Supporting Information), but also on whether or not the spine contains an inward (Type I) of outward drift (Type II).  An inward membrane drift, facilitating diffusion of membrane proteins into the spine was reported to be dependent on dynamin GTPase activity\cite{Henley}. Dynamin GTPase activity produces a membrane drift facilitating the transport of membrane material from  shaft to  spine.
%%%%%%%%%%%%%%%%%%%%%%%%%%%%%%%%%%%%%%%%%%%%%%%%%%%%%%
\subsection{Regulation of fluxes in dendritic spines}
%%%%%%%%%%%%%%%%%%%%%%%%%%%%%%%%%%%%%%%%%%%%%%%%%
Neuronal dendrites are organized in nanodomains that control the trafficking of key  excitatory and inhibitory receptors \cite{HozePNAS,DeepakJNS2013,SpechtTriller,Masson}. We showed here that some of these domains can be stable for a time scale of an hour and others can disappear at a time scale of tens of minutes.  These nanodomains can be structured by potential wells that reflect the interaction of receptors with molecular partners.  We reported here an example where a potential well regulates the flux of receptors at the base of a dendritic spine. A similar regulation restricting diffusion was previously reported \cite{Ashby} based on FRAP analysis: the regulation was described as imposed by a physical barrier at the neck of mature mushroom spines. We attributed here the restriction as a transient potential barrier possibly generated by local membrane reorganization due to molecular rearrangement. Such regulations make the two-dimensional spine surface a chemical compartment that can be isolated from the rest of the dendrite \cite{Yuste-book}. Interestingly, potential wells can appear either as isolated structures at the PSD inside a synapse \cite{HozePNAS}, or in clusters in the dendrite as shown in Figure \ref{figtimelapse}.  These results open new questions about the nature, the dynamics and the regulation of the potential wells over time scale ranging from minutes to hours.\\
The number of receptors inside the spine is a key component of the synaptic strength \cite{Bredt1}, which is independent of the receptor subtypes \cite{nicoll2013Nature}. This number depends on the residence time of moving receptors \cite{Holcman-triller,taflia}. In the two types of synapses (I and II), the residence time is quite different and controlled by different mechanisms (drift in opposite directions). Interestingly, the nature of the spines can change at a time scale of tens of minutes and a spine of type I can become type II or the converse \cite{HozePNAS}.\\
The method developed here could in principle be applied to other areas such as virology or any field where the amount of recorded trajectories is large enough. The method could also be applied for three-dimensional cellular domains. However, additional information would be needed to reconstruct the three dimensional domain. \\

%%%%%%%%%%%%%%%%%%%%%%%%%%%%%%%%%%%%%%%%%%%%%%%%%%%%%%
\subsection*{Supporting Information}\label{MM}
%%%%%%%%%%%%%%%%%%%%%%%%%%%%%%%%%%%%%%%%%%%%%%%%%%%%%%
 We developed a software in Matlab (MathWorks) to analyze the data and to simulate stochastic trajectories on empirical domains. A detailed description of our reconstruction algorithm is described in the Methods section.

%%%%%%%%%%%%%%%%%%%%%%%%%%%%%%%%%%%%%%%%%%%%%%%%%%%%%%
\subsection*{Acknowledgements}
%%%%%%%%%%%%%%%%%%%%%%%%%%%%%%%%%%%%%%%%%%%%%%%%%%%%%%
We thank D. Nair, J.B. Sibarita,  D. Chiquet and E. Hosy for fruitful discussions on this subject and for providing us with the experimental recording of trajectories. N. Hoze is supported by a Labex MemoLife fellowship.

%%%%%%%%%%%%%%%%%%%%%%%%%%%%%%%%%%%%%%%%%%%%%%%%%%%%%%%%%%%%%%%%%%%%%%%%%%%%%%%%%%%%%%%%%%%%%%%%%%%%%%%%%%%%

\newpage
%%%%%%%%%%%%%%%%%%%%%%%%%%%%%%%%%%%%%%%%%%%%%%%%%%%%%%%%%%%%%%%%%%%%
%\section*{Acknowledgments}
%%%%%%%%%%%%%%%%%%%%%%%%%%%%%%%%%%%%%%%%%%%%%%%%%%%%%%%%%%%%%%%%%%%%

%%%%%%%%%%%%%%%%%%%%%%%%%%%%%%%%%%%%%%%%%%%%%%%%
%%%%%%%%%%%%%%%%%%%%%%%%%%%%%%%%%%%%%%%%%%%%%%%%
\clearpage
%%%%%%%%%%%%%%%%%%%%%%%%%%%%%%%%%%%%%%%%%%%%%%%%%
%%%%%%%%%%%%%%%%%%%%%%%%%%%%%%%%%%%%%%%%%%%%%%%%%
\clearpage

%%%%%%%%%%%%%%%%%%%%%%%%%%%%%%%%%%%%%%%%%%%%%%%%
%%%%%%%%%%%%%%%%%%%%%%%%%%%%%%%%%%%%%%%%%%%%%%%%
\clearpage

%%%%%%%%%%%%%%%%%%%%%%%%%%%%%%%%%%%%%%%%%%%%%%%%
%%%%%%%%%%%%%%%%%%%%%%%%%%%%%%%%%%%%%%%%%%%%%%%%

%\clearpage

%%%%%%%%%%%%%%%%%%%%%%%%%%%%%%%%%%%%%%%%%%%%%%%%
%%%%%%%%%%%%%%%%%%%%%%%%%%%%%%%%%%%%%%%%%%%%%%%%

\clearpage

\end{document}